\documentclass[12pt, a4paper]{article}
\usepackage{graphicx}
\usepackage{amssymb}
\usepackage{amsmath}
\usepackage{bm}
\usepackage{color}
\usepackage{cite}
\usepackage{epsfig}

\newcommand{\beq}{\begin{equation}}
\newcommand{\eeq}{\end{equation}}
\newcommand{\ov}{\overline}

\usepackage[colorlinks=true, linkcolor=black, citecolor=black,
urlcolor=black]{hyperref}

\begin{document}

\begin{titlepage}

\begin{flushright}

KUNS-2570

\end{flushright}

\vskip 2cm
\begin{center}

{\Large
{\bf 
Survey of Higgs interpretations 
\\ 
of the diboson excesses
}
}

\vskip 2cm

Yuji Omura$^{1}$, Kazuhiro Tobe$^{1,2}$, and Koji Tsumura$^{3}$

\vskip 0.5cm

{\it $^1$
Kobayashi-Maskawa Institute for the Origin of Particles and the
Universe, \\ Nagoya University, Nagoya 464-8602, Japan}\\[3pt]
{\it $^2$Department of Physics,
Nagoya University, Nagoya 464-8602, Japan}\\[3pt]
{\it $^{3}$Department of Physics, Kyoto University, Kyoto 606-8502, Japan}\\ [3pt]

\vskip 1.5cm

\begin{abstract}
We investigate diboson signals in the Standard Model (SM) with an extended Higgs sector, 
motivated by the excesses in the diboson channels at the LHC. 
We begin with the unitarity sum-rules of the weak gauge boson scattering 
assuming the Higgs sector is extended. 
According to the sum-rules, 
we discuss the Higgs interpretations of the diboson signals and 
the consistency with the ATLAS diboson anomaly and other experimental constraints. 
As a concrete example, we propose a two-Higgs-doublet model where
the Yukawa coupling of an extra CP-even scalar with up-type quark is relatively large. 
The diboson ($WW$ and $ZZ$) signals can be explained by 2 TeV CP-even Higgs boson, 
while the partners, the CP-odd and the charged Higgs bosons in the extra doublet, 
are severely constrained by both the LHC direct search and 
the indirect search via flavor observables. 
Especially, in order to avoid the diboson resonance search in the $Vh$ channel, 
further extensions of the model are required. 
The diboson excess is correlated with the SM Higgs signals in our framework, 
so that the precise measurement of the SM Higgs boson is also important to test the Higgs interpretation.

\end{abstract}

\end{center}
\end{titlepage}

\section{Introduction}
\label{sec:intro}

In the year 2012, the LHC experiment finally discovered the Higgs particle which plays 
an crucial role in the electroweak symmetry breaking in the Standard Model 
(SM) \cite{Aad:2012tfa,Chatrchyan:2012ufa}. 
The observed data
still have large uncertainties, but it is consistent with the SM predictions.
We expect that the SM is not the complete theory of our universe, 
but we could conclude that the SM very well describes our nature up to the electroweak scale. 
The LHC has already started seeking new phenomena at $13$ TeV, 
and we have to think about what the next discovery is.

On the other hand, the hint for the new physics beyond the SM might be already shown 
by the results at the LHC Run-I.
In fact, several deviations from the SM predictions are reported at the $\sim 3\sigma$ level, 
and it would be possible to interpret them as the signals of unknown particles. 
Especially, we would like to focus on the excesses reported by the ATLAS collaboration
in the diboson resonance searches at the LHC with $\sqrt{s}=8$ TeV and 20.3~fb$^{-1}$ \cite{Aad:2015owa}.
The analysis has been done in the processes, $pp \to WZ$, $WW$, and $ZZ$,
where the weak bosons decay hadronically. Constructing the invariant
masses, the ATLAS has found the narrow resonances around $2$~TeV in the $WZ$, $WW$, and $ZZ$
channels with local significance of 3.4$\sigma$, 2.6$\sigma$, and 2.9$\sigma$, respectively \cite{Aad:2015owa}. 
Interestingly, the CMS collaboration also reports the moderate excesses in the diboson channels 
around the mass region \cite{CMS1,CMS2}.
Although it is too early to conclude that it is the discovery of new heavy particles, 
it is worthwhile to survey and study any possibilities of the resonances. 

As the candidates for the excesses, spin-1 vector bosons have been considered in
the frameworks of composite vector models \cite{Matsuzaki,Thamm:2015csa,Carmona:2015xaa,Fukano:2015uga} 
and $W'/Z'$ models \cite{Hisano:2015gna,Cheung:2015nha,Dobrescu:2015qna,Alves:2015mua,Gao:2015irw,Brehmer:2015cia,Cao:2015lia,Abe:2015jra,Abe:2015uaa}.\footnote{Triboson interpretation is given in Ref. \cite{Aguilar-Saavedra:2015rna}. 
General analyses have been done in Refs. \cite{Franzosi:2015zra,Cacciapaglia:2015eea,Allanach:2015hba}. 
Spin-0 candidates are also proposed recently \cite{Chiang:2015lqa,Cacciapaglia:2015nga,Sanz:2015zha}. }
The extra boson should have a mass of 2 TeV and a narrow width, i.e. less than $100$ GeV.  
The excesses seem to require the cross section of about $10$ fb for the diboson production, and 
then it will be difficult to evade the stringent bound from the two-lepton resonance search 
in the quark/lepton universal model \cite{DrellYanBound,DrellYanBound2}. 
For instance, so-called leptophobic $Z'$ model, where the SM leptons are not charged under $U(1)'$, 
has been proposed in Ref. \cite{Hisano:2015gna}, and the large production cross section is achieved,
although there is a tension with the recent result of $t \ov{t}$ resonance search at the CMS \cite{Khachatryan:2015sma}.

In this paper, we study the possibility that a spin-0 particle with $2$ TeV mass reproduces 
the diboson signals reported by the ATLAS collaboration \cite{Aad:2015owa}.  
Such a particle is, for example, predicted by the extended SM with
an extra scalar $SU(2)_L$ doublet. 
In our model, we introduce another $SU(2)_L$-doublet scalar which 
couples to the SM fermions. The extra scalar develops nonzero vacuum expectation value (VEV), and then
contributes to the electroweak symmetry breaking. The scalar is decomposed into CP-odd and CP-even neutral scalars
together with a pair of charged scalars. The Nambu-Goldstone bosons eaten by $W$ and $Z$ are given by the
linear combinations of the extra scalars and the ones from the SM Higgs doublet. 
After the symmetry breaking, two massive CP-even scalars appear and one of them should have
about 125 GeV mass corresponding to the Higgs discovery at the LHC. 
The other scalar could have 2 TeV mass and may explain the diboson excesses according to
the $WW$ and $ZZ$ couplings generated by the mixing between the two neutral scalars.
Unlike the $Z'$ particle, we may suffer from the production of the extra scalar, because
Yukawa couplings of Higgs bosons with the SM fermions are expected to be hierarchical 
corresponding to the mass hierarchy.  
The $(u, \,u)$ and $(d, \,d)$ elements of the Yukawa couplings may be too small to produce 
the sufficient large number of the extra scalars. 
This is, however, not so simple and would be a non-trivial question.
In general, two Higgs $SU(2)_L$ doublets independently couple with the SM fermions,
if they are not distinguished by a symmetry. The scalars, in fact,
could have large Yukawa couplings regardless of the fermion mass hierarchy. 
Some of the couplings are constrained by flavor physics, but the other still have large room to
be investigated at experiments. Especially, we consider the scenario that the  $(u, \,u)$ coupling with
the extra scalar is sizable, and discuss not only the diboson excess but also the impact on
the other observables: the SM Higgs signals and flavor physics.
In particular, the Higgs signals around 125 GeV are correlated with the diboson excesses,
because the couplings of the two CP-even scalars are related each other. 
Besides, we could find sum-rules for general extended Higgs 
models, so that it is possible to derive some model-independent predictions.

First of all, we investigate the Higgs interpretation of the diboson signal in the SM with 
extended Higgs sector.
In Sec. \ref{sec:section2}, we develop the generic analysis of the unitarity sum-rules 
for weak gauge boson scattering amplitudes, 
and discuss the consistency with the diboson excesses.
In Sec. \ref{sec:section3}, we introduce a two-Higgs doublet model (2HDM) as a concrete example, 
classified as type-III 2HDM. 
Then, the diboson signals produced by the extra neutral scalars are studied. 
We discuss the impact on the 125 GeV Higgs signals as well as flavor constraints, in Sec. \ref{sec:section3}.
Our study is summarized in Sec. \ref{sec:summary}.

\section{Higgs Interpretations of the Diboson Signal}
\label{sec:section2}

Since we are interested in the Higgs interpretation of the diboson signal, 
a new Higgs boson, $H$, must couple to the weak gauge bosons. 
Such interactions give additional contributions to the weak boson
scattering amplitude.
In the SM, the high energy behavior of the weak boson scattering
amplitude is unitarized
by the Higgs boson contribution.
If the Higgs sector is extended to have extra neutral Higgs bosons,
the unitarity requirement of the amplitude leads sum-rules among the
Higgs couplings.
From the $VV\to VV$ amplitude with arbitrary number of the neutral Higgs bosons\footnote{
To be more precise, charged Higgs bosons can also be introduced in this setup 
if there is no $\phi^\pm_n W^\mp Z$ interaction term as in multi-Higgs-doublet models. 
},
we have
\begin{align}
\sum_{n=1}^N (\kappa_V^{\phi_n})^2 = 1,
\end{align}
where $\kappa_V^{\phi_n}\, (n=1,2,3, ..., N)$ are the coefficients of
the $\phi_nVV$ interaction term
normalized by the SM Higgs coupling \cite{Gunion:1990kf,Grinstein:2013fia,Nagai:2014cua}.
Similarly, the unitarization of the $VV\to f\bar f$ process gives
\begin{align}
\sum_{n=1}^N \kappa_V^{\phi_n}\, \kappa_F^{\phi_n} = 1,
\end{align}
where $\kappa_F^{\phi_n}$ are the normalized Yukawa coupling constants \cite{Gunion:1990kf}.
\\

Let us focus on the case with $N=2$. 
We call $\phi_1=h$, which is identified as the $125$~GeV Higgs boson,
and $\phi_2=H$, which will be regarded as the reported diboson resonance.
In this case, the Higgs-gauge couplings are parameterized by one
mixing angle $\theta$ as
\begin{align}
\kappa_V^h=\sin\theta, \qquad \kappa_V^H=\cos\theta,
\label{Eq:KV}
\end{align}
in order to satisfy the unitarity sum-rule derived from the vector
boson scattering.
The scaled Yukawa couplings are also parameterized by introducing 
one more parameter $\xi_F^{}$ as
\begin{align}
\kappa_F^h=\sin\theta-\xi_F^{}\,\cos\theta, \qquad
\kappa_F^H=\cos\theta+\xi_F^{}\,\sin\theta.
\label{Eq:KF}
\end{align}
Note that $\xi_F^{}$ can be a flavor-dependent parameter.
For instance, $\theta=\alpha$ and $\xi_F^{}=0$ are predicted in the real singlet extension 
of the Higgs sector, where $\alpha$ is the mixing angle between 
the SM-like Higgs boson and the singlet. 

We will see below that the required $\kappa_F^H$ (or $\xi_F^{}$) is about order of $10^4$ for $F=u$. 
In order to realize such large $\xi_F^{}$, only the possibility would be the multi-doublet structure 
of the Higgs sector because of allowed Yukawa interactions in renormalizable theories.  
The minimal model is 2HDM, where the two Higgs doublets develop nonzero VEVs.
In the 2HDM, $\theta=\beta-\alpha$ is related to 
the VEV mixing $\beta$ and CP even Higgs mixing $\alpha$, 
and $\xi_F^{}$ depends on $\tan\beta$ and extra Yukawa couplings. 
\\

Next, we discuss the diboson signals in this simple framework. 
The ATLAS collaboration has reported the excesses in the diboson resonance search
at the LHC with $\sqrt{s}=8$ TeV and 20.3~fb$^{-1}$ \cite{Aad:2015owa}.
The excesses are shown in the channels with the boson-tagged fat jets. 
The event number is around $10$, with about $0.1$ efficiency,
so that the production cross section for the diboson should be ${\cal O}(10)$ fb.
Moreover, the ATLAS distinguishes the $WZ$, $WW$ and $ZZ$ final states and
found the narrow resonances around $2$~TeV in the each
channels with local significance of 3.4$\sigma$, 2.6$\sigma$, and 2.9$\sigma$, respectively. 
It is impossible for one 2-TeV particle to reproduce the excesses in the all channels, because
of the charge conservation. In fact, the discrimination is done allowing $20$\% overlapping of signal events,
so that we could not conclude that the narrow resonances should be realized in the each final states separately. 
\\

For the moment, we assume the coupling relations given in Eqs.\eqref{Eq:KV} and \eqref{Eq:KF} 
without specifying the explicit form of the Higgs sector. 
If the new Higgs boson $H$ is produced from the $u\bar u$ pair annihilation 
and decays into weak gauge boson pairs, 
the diboson signal rate is determined by $\kappa_V^H(\equiv \xi_V^{})$ and $\kappa_u^H$ (or $\xi_u$). 
\footnote{
The $d\bar d$ pair annihilation may also be used for $H$ production if $\kappa_d^H$ is large. 
However, it is not efficient at the LHC because of the parton distribution in protons. 
Therefore, we here concentrate on the production from $u\bar u$. 
}
\footnote{
We give a brief comment on the charged Higgs interpretation of the diboson excess, 
where the charged Higgs boson $H^\pm$ decaying to $WZ$. 
Such charged Higgs bosons can in principle appear in models with 
a triplet \cite{Konetschny:1977bn,Magg:1980ut,Cheng:1980qt,Schechter:1980gr} 
or more higher isospin multiplets with a VEV \cite{Ren:2011mh,Hisano:2013sn,Kanemura:2013mc}. 
However, we cannot realize the sufficiently large production cross sections for $H^\pm$. 
Because of nonzero electric charge of $H^\pm$, the gluon-gluon fusion (GGF) mechanism cannot work. 
The vector boson fusion (VBF) is less significant, because of the stringent constraint from the electroweak 
precision data on the VEV. 
Thus, the $q\bar q'$ annihilation through the Yukawa coupling is only the possibility for producing $H^\pm$ 
similarly to the neutral Higgs boson case, which can only be realized from the doublet Higgs field. 
}
\\

\begin{figure}[tbh]
\centering
\includegraphics[height=7cm]{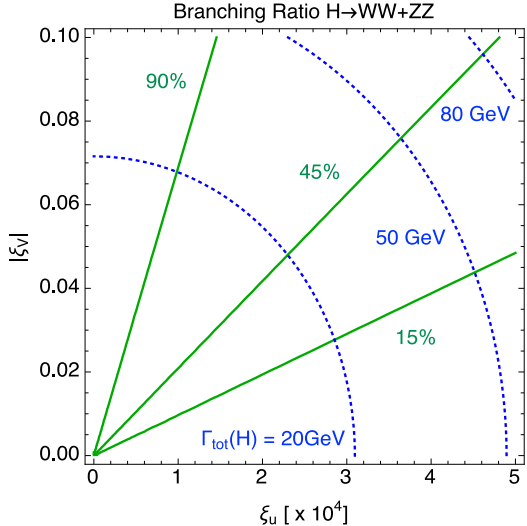}
\caption{The branching ratios, ${\cal B} (H \to ZZ+ WW)$, (green solid lines) and the total width of $H$, $\Gamma_{\rm tot}(H)$, (blue dotted curves) are plotted 
as functions of $\xi_u$ and $|\xi_V^{}|$, assuming $\xi_{F\neq u}=0$. $\xi_u$ is fixed to ${\cal O} (10^{4})$, which corresponds to ${\cal O} (0.1)$ Yukawa coupling in 2HDM. 
}
\label{Fig:BR-VV}
\end{figure}
In Fig.~\ref{Fig:BR-VV}, 
the branching fractions, ${\cal B} (H \to ZZ+ WW)$, with $M_H=2$ TeV are shown (green solid lines) as the function of these two parameters, $\xi_u$ and $|\xi_V^{}|$. $\xi_{F\neq u}=0$ is assumed and $\xi_u$ is focused on the ${\cal O} (10^{4})$ range, which corresponds to ${\cal O} (0.1)$ Yukawa coupling in 2HDM. 
Because there can be contamination of the diboson signal events $(WW, WZ, ZZ)$, the diboson channels are 
simply summed. The ratio of $WW:ZZ$ decay channel is almost $2:1$. 
We have also plotted the total width of $H$ in Fig.~\ref{Fig:BR-VV} (blue dotted curves). 
The total width of the resonance is still uncertain due to the detector resolution. 
The experimentally allowed width is smaller than about 100 GeV, 
so that most of the parameter space in this plot satisfies this loose constraint.  
\begin{figure}[tbh]
\centering
\includegraphics[height=7cm]{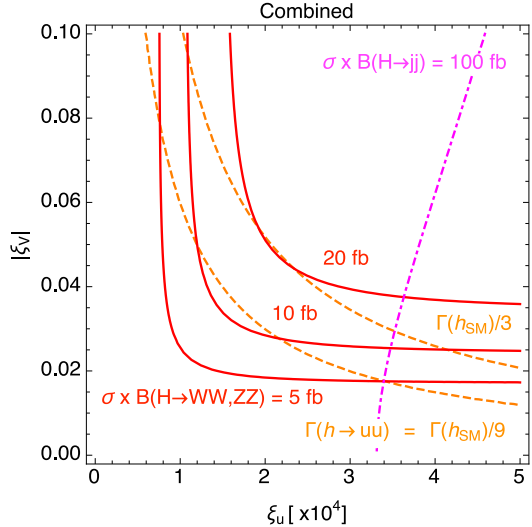}
\caption{The extra Higgs boson interpretation of the diboson excess (red solid curves). 
The bound from the dijet resonance search is also plotted (the magenta curve). 
The anomalous width of the 125 GeV Higgs boson decaying to $u\bar u$ is also shown (yellow dashed curves). 
}
\label{Fig:signal}
\end{figure}
\\

In Fig.~\ref{Fig:signal}, we show the diboson signal rates, $\sigma(pp \to H) \times {\cal B} (H \to ZZ+ WW)$ (red solid curves). 
Using the MadGraph \cite{madgraph}, the leading order (LO) production cross section of 2-TeV $H$ 
at the LHC with the collision energy of 8 TeV is estimated as
\beq
\sigma (pp [u\bar u] \to H) \simeq \left ( \sqrt 2  M_u\, \kappa^H_u/v \right )^2 \times 521 \, [ {\rm fb}],
\eeq
where $M_u=2.3$ MeV is taken. 
The total width of $H$ is assumed to be $50$ GeV in the calculation of the Higgs production, $\sigma(pp \to H)$, but it is insensitive to the production cross section. 
There can be large QCD corrections to the signal process. 
The production cross sections via
$\overline{u}u\to Hg$ (real gluon emissions) and $gu\to uH$ are $370\,$fb and $57\,$fb 
normalized by $(\sqrt 2  M_u\, \kappa^H_u/v)^2$, 
which are not included in the signal cross sections. 
The effect of the high order correction may relax the allowed parameter space.
The dijet decay channel of $H$ also give a constraint such like 
$\sigma(pp \to H) \times {\mathcal B}(H\to jj) < 100$ fb \cite{CMSdijet}. 
The right-side region of the majenta dot-dashed curve in~Fig. \ref{Fig:signal} is excluded. 
In addition, we present the anomalous partial width of $h\to u\bar u$ (yellow dashed curves), 
which is negligible in the SM. 
In our setup, the Yukawa coupling of the 125 GeV Higgs boson is related by Eq. \eqref{Eq:KF}, 
so that the large Yukawa coupling of $H$ induces the anomalous decay of $h$. 
For $\Gamma(h\to u\bar u)=\Gamma(h_\text{SM}^{})/9$ $(\Gamma(h_\text{SM}^{})/3)$, 
10\% (25\%) of the produced 125 GeV Higgs boson decays into $u\bar u$ in this case. 
\\

\begin{figure}[tbh]
\centering
\includegraphics[height=7cm]{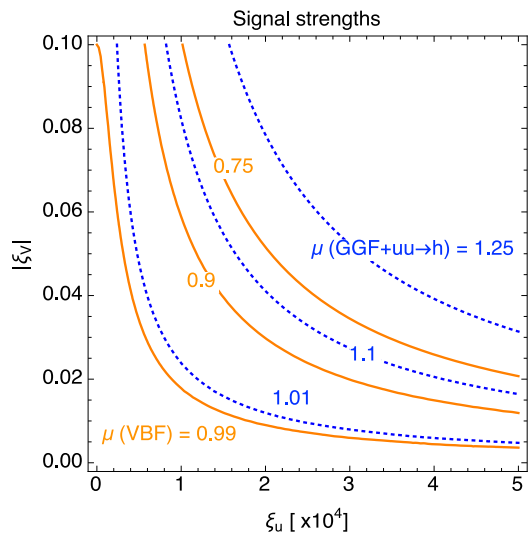}
\caption{The modified signal strength of the 125 GeV Higgs boson in 
the VBF (yellow solid curves) and the GGF (blue dashed curves). 
The contribution of the anomalous production from $u\bar u$ annihilation is added to the GGF process. 
}
\label{Fig:mu}
\end{figure}

We also estimate the modification of the signal strengths of the 125-GeV Higgs boson in each production mechanism in Fig.~\ref{Fig:mu}. 
The signal strengths are corrected by two parts. For the VBF, the production rate 
is slightly changed by $\kappa_V^h(=\sqrt{1-\xi_V^2})$, while the decay part is modified as in Fig.~\ref{Fig:signal}. 
The signal strength of the VBF mechanism is given by yellow solid curves, which are almost the same as 
the yellow dashed curves in Fig.~\ref{Fig:signal} due to the smallness of the $\xi_V^2$. 
($\mu$(VBF)$=0.9, \,0.75$ correspond to $\Gamma(h\to u\bar u)=\Gamma(h_\text{SM}^{})/9, \, \Gamma(h_\text{SM}^{})/3$ respectively.) 
On the other hand, the signal strength of the GGF production would 
be contaminated by the anomalous production mechanism through $u\bar u$ fusion. 
Thus, we evaluated the sum of these two production mechanisms for calculating the signal strength. 
The anomalous production cross section at LO is calculated by MadGraph as
\beq
\sigma (pp[u\bar u] \to h) \simeq \left ( \sqrt2 M_u\, \kappa^h_u/v \right )^2 \times 54 \, [ {\rm nb}]. 
\eeq
The cross section of the conventional GGF is $19.27$ pb at the NNLL QCD and the NLO EW. 
In total, the signal strength from the GGF is enhanced by the additional anomalous contributions. 
The impact on the phenomenology of the 125 GeV Higgs boson is the characteristics of 
the extra Higgs interpretation of the diboson excess.

\section{A Renormalizable Model}
\label{sec:section3}
In this section, we give a concrete example of our analysis presented above. 
Let us simply add one extra Higgs $SU(2)_L$-doublet scalar to the SM. 
If they are not distinguished by
imposing symmetry on the doublets, both can couple with the SM fermions.
The two Higgs doublets may develop the nonzero VEVs, but we can take the base
that only one Higgs doublet develops nonzero VEV and the other does not, without loss of generality.
In such a base, the Yukawa coupling can be described as
\begin{align}
  {\cal L}&=-\bar{Q}_L^i H_1 y^i_d d_R^i -\bar{Q}_L^i H_2 \rho^{ij}_d d_R^j 
-\bar{Q}_L^i (V^\dagger_{\rm CKM})^{ij}\tilde{H}_1 y^j_u u_R^j -\bar{Q}_L^i (V^\dagger_{\rm CKM})^{ij}\tilde{H}_2
  \rho^{jk}_u u_R^k 
\end{align}
where $Q=(V_{\rm CKM}^\dagger u_L,d_L)^T$, 
$V_{\rm CKM}$ 
is the Cabbibo-Kobayashi-Maskawa 
matrix
and the fermions $(f_L^{},~f_R^{})$ $(f=u,~d
)$ are mass eigenstates.
Here, $H_1$ and $H_2$ are defined as the scalars decomposed as follows: 
\begin{eqnarray}
  H_1 =\left(
  \begin{array}{c}
    G^+\\
    \frac{v+\phi_1+i\,G^0}{\sqrt{2}}
  \end{array}
  \right),~~~
  H_2=\left(
  \begin{array}{c}
    H^+\\
    \frac{\phi_2+i\,A}{\sqrt{2}}
  \end{array}
  \right),
\end{eqnarray}
where $G^+$ and $G^0$ are the Nambu-Goldstone bosons eaten by $W$ and $Z$ bosons, 
and $H^\pm$ and $A$ are charged Higgs bosons and a CP-odd scalar boson, respectively. 
The VEV, $v$, which gives weak gauge boson and fermions masses, is fixed to $v \simeq 246$ GeV.
The CP-even neutral Higgs bosons, $\phi_1$ and $\phi_2$, mix each other and the mass 
eigenstates, $h$ and $H$ ($M_H>M_h$), are given by 
\begin{eqnarray}
  \left(
  \begin{array}{c}
    \phi_1\\
    \phi_2
  \end{array}
  \right)=\left(
  \begin{array}{cc}
    \cos(\beta - \alpha) & \sin(\beta- \alpha)\\
    -\sin (\beta- \alpha) & \cos(\beta -\alpha)
  \end{array}
  \right)\left(
  \begin{array}{c}
    H\\
    h
  \end{array}
  \right).
\end{eqnarray}
In our argument, $M_h  =125$ GeV and $M_H=2$ TeV are assumed. 
In the mass eigenstate of scalars, the Yukawa interactions are expressed as
\begin{align}
  {\cal L}&=-\sum_{\phi=h,H,A} y_{\phi i j}^{}\bar{f}_{Li}^{} \phi f_{Rj}^{}
-\bar{u}_i(V_{\rm CKM}\rho_d P_R-\rho_u^\dagger V_{\rm CKM} P_L)^{ij} H^+d_j+{\rm h.c.},
\end{align}
where
\begin{align}
  y_{hij}^{}&=\frac{M_{f}^i}{v}s_{\beta\alpha}\delta_{ij}+\frac{\rho_{f}^{ij}}{\sqrt{2}} c_{\beta\alpha},\quad
  y_{Hij}^{}=\frac{M_f^i}{v} c_{\beta \alpha}\delta_{ij}-\frac{\rho_f^{ij}}{\sqrt{2}} s_{\beta\alpha},\quad 
  y_{Aij}^{}=
    -\frac{i\rho_f^{ij}}{\sqrt{2}}\,(2T_{3f}^{}),
\label{Yukawa}
\end{align}
where $s_{\beta \alpha}=\sin (\beta - \alpha)$, $c_{\beta \alpha}=\cos(\beta -\alpha)$ and  
$M^i_{f}$ are the mass engenvalules of the SM fermions. 
The (aligned) SM limit corresponds to $s_{\beta \alpha} \to 1$ in this expression. 
The anomalous Yukawa matrices $\rho_f^{ij}$ are general 3-by-3 complex matrices, 
so that the off-diagonal elements generally induce tree-level Flavor Changing Neutral Currents (FCNCs) via the SM Higgs boson, 
unless $s_{\beta \alpha}=1$ is satisfied. 
How to control the FCNCs is beyond our scope, but we simply assume that
$\rho_f^{ij}$ is in the diagonal form: $\rho_f^{ij}=\rho_f^{i} \delta^{ij}$. 
We are especially interested in the $(u, \,u)$ element of $\rho_u^{ij}$.
In order to enhance the diboson production at the LHC,
we will find that $\rho_u^{u} \sim {\cal O}(0.1)$ is required, which is unusually larger than the typical value
suppressed by the up quark mass, as we have already discussed in Sec. \ref{sec:section2}. 
The magnitude of $\xi_u\sim{\cal O}(10^4)$ corresponds to $\rho_u^{u} \sim {\cal O}(0.1)$ in this model.
We assume that the other diagonal components, $\rho^i_f$, are negligibly small.
\\

\begin{figure}[tbh]
\centering
\includegraphics[height=7cm]{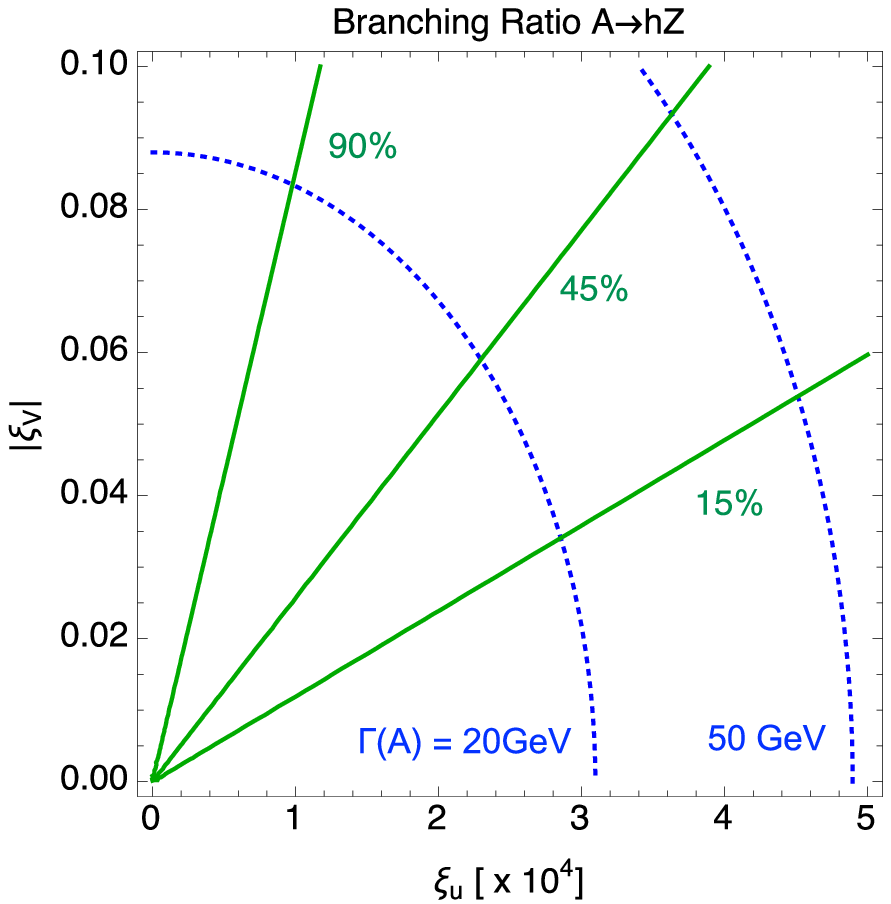}
\includegraphics[height=7cm]{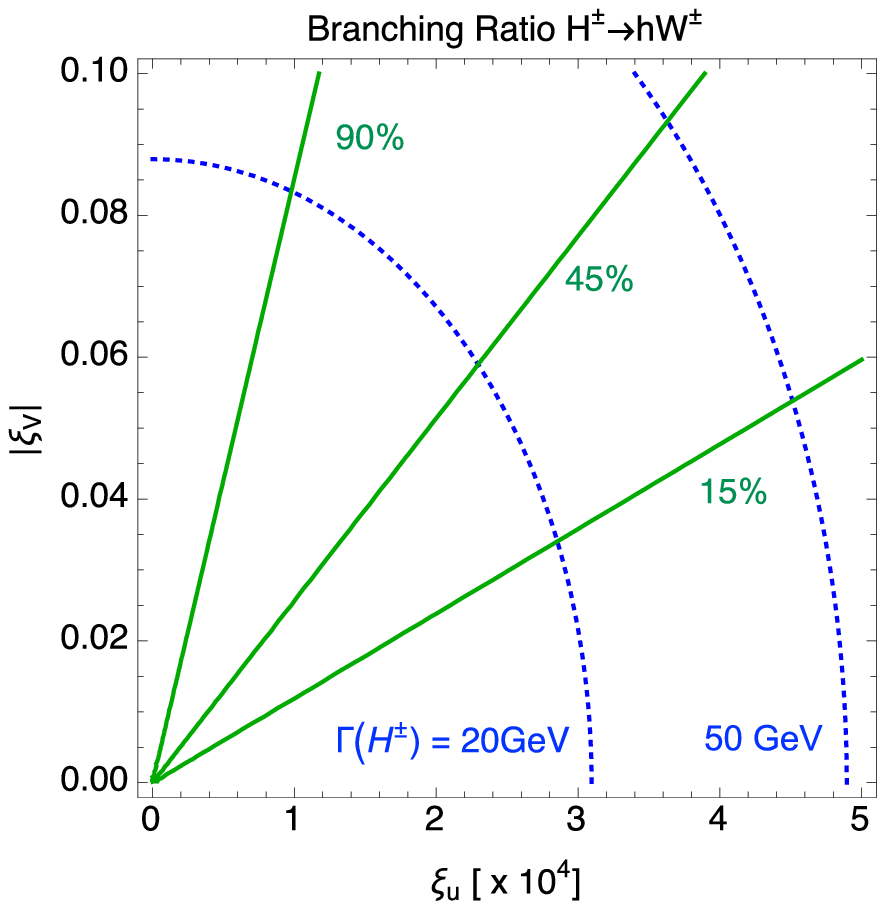}
\caption{The bosonic decay branching ratios, ${\cal B} (A \to h Z)$ and  ${\cal B} (H^\pm \to h W^\pm)$, (green solid lines) of the CP odd Higgs boson (left) and 
the charged Higgs boson (right) are shown. 
The total widths of $A$ and $H^\pm$ ($\Gamma(A)$ and $\Gamma (H^\pm)$) are also given by the blue dotted curves. 
}
\label{Fig:decays}
\end{figure}

The massive charged Higgs boson ($H^{\pm}$) and the CP-odd Higgs boson ($A$) generally have 
different masses from the CP-even masses, but the mass differences should be small in order to avoid 
the conflict with the electroweak precision observables and 
the perturbative unitarity constraint \cite{Kanemura:1993hm,Akeroyd:2000wc}.
\footnote{Note that this situation is naturally realized when the extra scalars are heavy, because 
the mass differences are roughly scaled as $\Delta m \simeq (\lambda_i-\lambda_j) v^2/M$, using dimensionless parameters, $\lambda_{i}$, in the Higgs potential  \cite{Kanemura:2011sj}.} 
The masses should be around 2 TeV as well, and then they may also contribute to the diboson signals   
or the diboson/dijet constraints as we see later. 
In Fig. \ref{Fig:decays}, each panel shows the branching ratios of $H^{\pm}$ and $A$ decaying to 
$h W^{\pm}$ and $h Z$, respectively. The rest of decay channels is assumed to the dijet modes. 
When $\xi_u$ is ${\cal O}(10^{4})$, the bosonic branching ratios, ${\cal B} (A \to h Z)$ and  ${\cal B} (H^\pm \to h W^\pm)$, are estimated as  ${\cal O}(10)$ \%. 
\\

\begin{figure}[tbh]
\centering
\includegraphics[height=7cm]{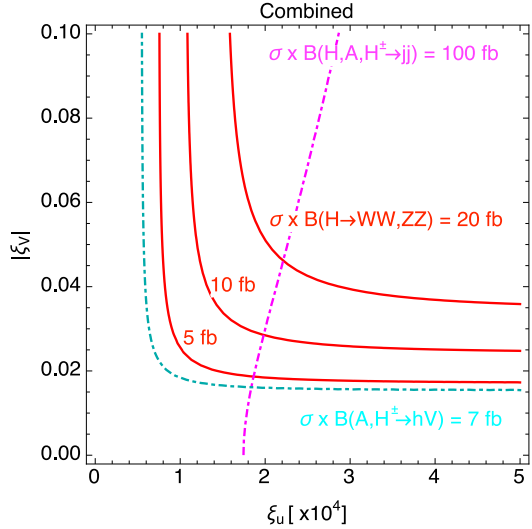}
\caption{Direct resonance search bounds from the dijet channel (the majenta dot-dashed curve) 
and from the $Vh$ channel (the cyan dot-dashed curve)
in the 2HDM. 
}
\label{Fig:2HDM-direct}
\end{figure}
\begin{figure}[tbh]
\centering
\includegraphics[height=7cm]{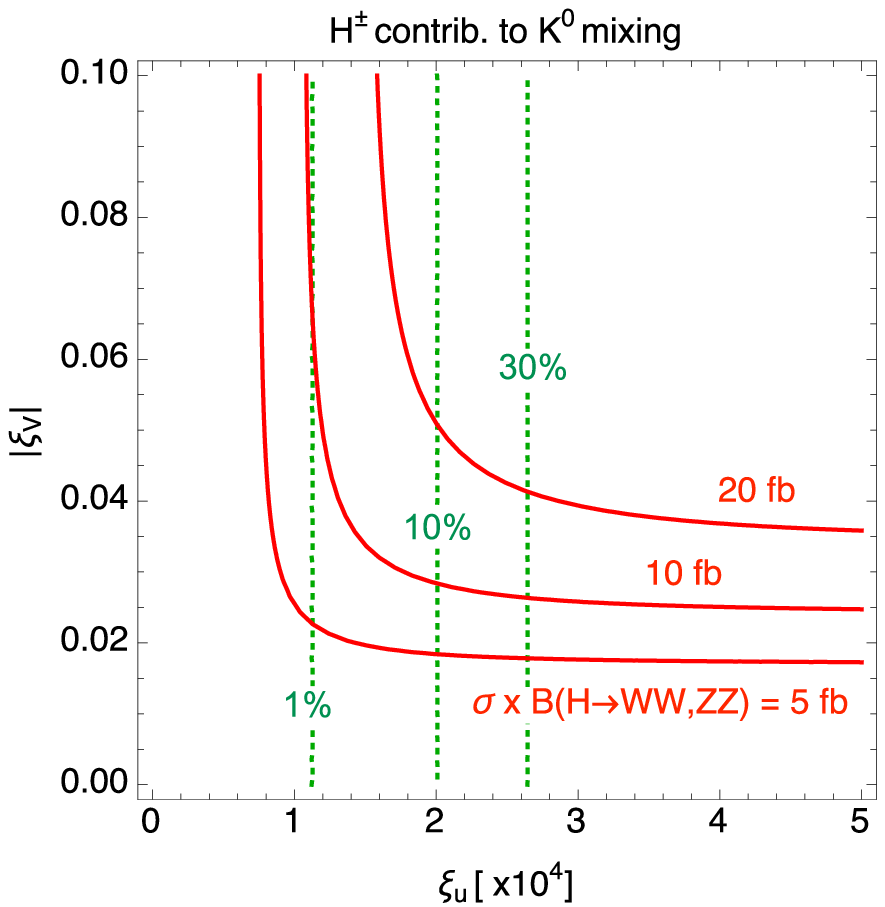}
\caption{The short distance contribution (green dashed lines) to the $K^0$-$\overline{K^0}$ mixing ($\Delta M_K$)
from the charged Higgs boson mediated box diagram, normalized by the SM contribution.
}
\label{Fig:K0-mixing}
\end{figure}
The CMS collaboration has reported an excess in the $h W^{\pm}$ channel with $l \nu b \ov{b}$ in the final state \cite{CMSWh}, but the recent analysis in hadronic final states shows strong constraints on the diboson production of a heavy particle ($V'$) associated with the 125-GeV Higgs: $\sigma (pp\to V')\times\text{BR} (V' \to V h) \lesssim 7$~fb ($V=W$, $Z$)\cite{Khachatryan:2015bma}. %
The production cross sections of $H^{\pm}$ and $A$ are the same order of the one of $H$, so that 
the $Vh$ resonance search would give a severe constraint to the diboson production of $H$. 
We have calculated the cross sections of $H^{\pm}$ and $A$ similarly to $H$ and evaluate the bound in Fig.~\ref{Fig:2HDM-direct}. 
Moreover, $H^{\pm}$ and $A$ decay to two jets according to the large $\xi_u$, and then the dijet bound
should be reanalyzed including the extra contributions. 
In Fig. \ref{Fig:2HDM-direct}, we correct the direct search bound in the 2HDM 
from the dijet decay mode provided by the sum of all extra scalars (majenta dot-dashed curve) 
and from the diboson decay mode of $A \to hZ$ and $H^\pm \to hW^\pm$ (cyan dot-dashed curve). 
Considering the bound from the $Vh$ resonance search, the $WW$ and $ZZ$ productions through $H$ are
strongly limited,   
as far as $H^{\pm}$ and $A$ do not decay into extra particles. 
Since $A$ and $H^\pm$ are the $SU(2)$ partners of $H$ in the SM limit, 
the production cross sections and the decay branching fractions are strongly correlated 
with the help of the equivalence theorem. 
\footnote{
In the case of the spin-1 interpretation, the equivalence of the weak gauge boson scattering amplitudes 
and the NGB scattering amplitude imply the relation between $VV$ and $Vh$ decay channels, and 
the constraint from $Vh$ resonance search cannot be evaded. 
}

Let us consider the possibility that $A$ and $H^\pm$ masses are degenerate but heavier than $M_H$. 
As we noted, the mass differences require relatively large quartic coupling constants, 
$\lambda_i \sim \Delta M\, M/v^2$, in the Higgs potential. 
For $\Delta M \simeq 100$ GeV, $\lambda_i$ should be order of unity. 
Such a large coupling results in the Landau pole at near TeV scale.  
If the mass difference is around $100$ GeV, the production cross sections of $A$ and $H^\pm$
becomes about 70 \%. 
Besides, the branching ratios of $A$ and $H^\pm$ to $H \, Z$ and $H \, W^\pm$ are about a few \%.
 Thus, the constraint from the $Vh$ resonance search would 
 require further extensions of the model. 
\\

We here show simple examples of such extensions which avoid the constraint from the $Vh$ resonance search. 
Let us introduce extra $SU(2)_L$-singlet scalars, $s^-$ and $s_0$,  with the hypercharge $-1$ and $0$ respectively. 
We also assign $Z_2$-odd charges to $s^-$ and $s_0$, where the other particles are $Z_2$-even.
Then, we could write the following quartic coupling in the Higgs potential:
\beq
\lambda_c \, s_0 s^- \epsilon_{ab} H^a_1H^b_2+\text{H.c.} =\lambda_c \frac{v}{\sqrt{2}} \, s_0  s^- H^+ + \dots 
\eeq
This terms gives the trilinear coupling, $s_0s^-H^+$, and $H^+$ can decay to $s_0$ and $s^-$,
if they are light enough to be allowed kinematically.

Similarly, we can introduce a term as 
\beq
\lambda_0 \, s^2_0  H^\dagger_1H_2+\text{H.c.} =-\text{Im}(\lambda_0) v \, s^2_0  A +\dots
\eeq
If the extra-decay branching ratios are dominant, we could evade the bound from the $Vh$ resonance search
at the CMS experiment \cite{Khachatryan:2015bma} and also from the dijet resonance search \cite{CMSdijet}. 

In order to avoid the reduction of the diboson signals of $H$, we have to tune the dimensionless parameters,
$\lambda_c$ and $\lambda_0$: for instance, the real part of $\lambda_0$ should be small compared with the imaginary part. If $s_0$ is lighter than $s^\pm$, $s^\pm$ can decay to $s_0$ and SM fermions through the 
heavy Higgs exchanging. $s_0$ is stable according to the $Z_2$ symmetry and could be a good dark matter candidate.
\\

Finally, we study the flavor constraint in our 2HDM. In our scenario, the only $(u, \, u)$ element of $\rho_u^i$ is 
large and the other elements are negligible.
The FCNC involving the neutral scalars are expected to be small, because of the imposed alignment.
The flavor changing Yukawa couplings of $H^{\pm}$ however may be sizable because of the CKM matrix
and the large $\rho_u^u$. The one-loop box diagram involving $H^{\pm}$ and up-type quark via the sizable
Yukawa couplings enhances $K^0$-$\overline{K^0}$ mixing, as we see in Fig. \ref{Fig:K0-mixing}.
Our signal region predicts the 1-10 \% modifications of the SM prediction, assuming $\rho_u^u$ is real.
Such a large new physics contribution to $\Delta M_K$ is still allowed because of the large uncertainty 
of the long distance effect \cite{Bona:2007vi}.

\section{Summary}
\label{sec:summary}

We consider the possibility that the spin-0 particle resides around 2 TeV mass region, motivated by the
excesses at the ATLAS experiment \cite{Aad:2015owa}.
These possible signals are found in the diboson resonance searches with
two fat jets in the final state, and requires $\sim 10$~fb cross
sections with narrow decay widths. Our spin-0 particle couples with the weak bosons and fermions 
by analogy to the SM Higgs particle. The production of the extra scalar requires a
relatively large value of the Yukawa coupling with light quarks, compared to the SM Higgs couplings.
We assume that the only $(u, \, \overline{u})$ coupling of the heavy scalar is sizable, and 
it decays to the weak bosons and the two quarks dominantly through the mixing with the 125-GeV Higgs boson.
We give a 
generic analysis in multi-Higgs models, and we conclude that
$\sigma$ ($pp \to H \to VV$) $ \simeq {\cal O}(10)$ fb is possible in the spin-0 scenario, without conflict with
the dijet resonance search.
It is important to investigate the consistency with the SM Higgs signal around 125 GeV as well.
The results on the 125-GeV Higgs are not still conclusive to test our model, but the diboson excesses 
predict the deviation from the SM prediction, so that the signal strengths of the SM Higgs are
considerably relevant to our model. In fact, the 10-25 \% deviation of the dijet production
from the 125-GeV Higgs boson limits the diboson productions: $\sigma$ ($pp \to H \to VV$) $  \lesssim 10$-$20$ fb.
In other word, the $10$ \% accuracy is required to test our model in the Higgs signals. 

Based on the generic analysis, we discussed an explicit 2HDM, where one extra Higgs doublet couples with
only up-type quark strongly in the mass eigenstate.  
In this model, the Heavy charged Higgs and pseudoscalar around 2 TeV are also predicted, and then 
$Vh$ resonance search gives the strongest bound. If the bound is taken into account, the $WW$ and $ZZ$ productions
cross sections are limited to about $5$ fb. This is the common issue in the extended SMs to explain the diboson anomaly.
Introducing extra scalars, we can evade the stringent bound without  reducing the diboson signals of $H$,
although we may tune some parameters in the scalar potential.
As for the flavor constraint, the strongest bound would be from the K system, because of the sizable CKM matrix elements involving the first and second generations. The impact of the diboson signals on the $K_0$-$\overline{K_0}$ mixing is at most O(10) \% correction to the SM, which is not excluded in our signal region at this present time.

The 2-TeV resonance decaying to diboson reported by the ATLAS collaboration can be explained by the extra Higgs boson in a renormalizable theory. A general impact on the 125-GeV Higgs boson from the diboson excesses is modification of the Higgs signal strengths, which will be measured more precisely at the future colliders. The direct searches for the partners of the 2 TeV resonance are very exciting at LHC Run-II.

\vspace{0.5cm}

\noindent{ {\it Note Added}}

While we are finalizing this paper, there appeared a paper, which similarly proposes a 2HDM, motivated by the ATLAS diboson excesses \cite{Chen:2015xql}. 
The production mechanisms for 2 TeV resonances are similar, but the different mass spectrum and decay chains
are studied.

\section*{Acknowledgments}
This work is supported by  the MEXT Grant-in-Aid for
Scientific Research on Innovative Areas, Japan, 
Nos. 23104011 (for Y. Omura), 26104705 (for K. Tobe), 
and 26104704 (for K. Tsumura). 


\appendix



\end{document}